# A Systematic Review of the Critical Factors for Success of Mobile Learning in Higher Education (university students' perspective)

Muasaad Alrasheedi, Luiz Fernando Capretz and Arif Raza
Department of Electrical and Computer Engineering, Western University
1151 Richmond St, London, Ontario, N6A 3K7, Canada
{malrash, lcapretz, araza22}@uwo.ca

## ABSTRACT

The phenomenon of the use of a mobile learning (m-Learning) platform in educational institutions is slowly gaining momentum. While this can be taken as an encouraging sign, the perplexing part is that the fervor with which mobile phones have been welcomed into every aspect of our lives does not seem to be evident in the educational sector. In order to understand the reason, it is important to understand user expectations of the system. This paper documents a systematic review of various research studies seeking to find the success factors for effective m-Learning. A total of 30 studies were included in the research, which combined would give a true picture of user perceptions of the factors they consider important for effective m-Learning implementation. Our systematic review collates results from 30 studies conducted in 17 countries, where 13 critical success factors (CSFs) were found to strongly impact m-Learning.

## INTRODUCTION

The idea of m-Learning, a relatively new concept, is made interesting by the way it blends the notion of mobility into the already popular electronic learning context. The concept of mobility actually makes the concept of m-Leaning even more revolutionary than electronic learning (Ally, 2009). Learner mobility as well as educator mobility not only removes the learners from the physical constraints of a particular learning location but also the constraints of learning time. In other words, a learner can control what they want to learn, when they want to learn, and where they want to learn. They are not restricted to prescribed materials, a physical classroom, or even a particular time around which they have to schedule other activities (Kukulska-Hulme, 2005).

Research in m-Learning has always been distributed and non-cohesive. One of the reasons is the inherent disagreement as to what constitutes m-Learning in the first place. As the specific definition used by a particular researcher automatically decides the scope of the research, the ensuing studies have been equally diversified in contexts and methodologies. Mobile learning was originally defined from a device-centric perspective. The most refined definition from this point of view was given in 2009 by (Traxler, 2009), who described the technology of m-Learning to include both software and hardware that enabled the learning devices to be portable. Device-based definitions, however, limit the scope of m-Learning, as m-Learning is not merely a conjugation of the words mobile and learning. Similarly, m-Learning is distinctly different from e-Learning and cannot be defined in the words of Traxler as "eLearning made mobile" (Traxler, 2009). The rapid changes in technology have also proved to be a hindrance to researchers attempting to define m-Learning in terms of devices.

One of the popular definitions encompasses the mobility and technological aspects, where m-Learning is characterized by its anytime-anywhere learning capacity and use of multiple media functions like pictures, videos, text, and voice (Shih & Mills, 2007). In addition to the unfettered nature of learning in terms of space and time, m-Learning additionally includes ideas like spontaneity, interactivity, informality, and ownership of learning (Traxler, 2008).

Basically mobile technology has seen high penetration in all aspects of people's lives, however its usage as an educational platform has been very slow. There are definite barriers to adoption of m-Learning platform, especially by higher learning institutions. Multiple researches have been conducted in various countries across the world to evaluate the success factors of mobile learning in higher education. The studies are fragmented and meta-analyses of the studies have focused on the geographical clusters, learner profiles and mobile device types. There is a need for a research that collates the studies in the area of m-Learning in terms of factors that users perceive to be important for success.

Firstly this study is a systematic review of existing studies to determine the critical success factors for m-Learning in higher education. Secondly an evaluation of studies is conducted in 17 countries, which means that the success factors can reasonably be expected to be universal. Thirdly, a mathematical evaluation of factors is also carried out using a common method and scale (Likert 5-point).

Finally, this study gives a comprehensive understanding of factors that learners (worldwide) expect in a good mobile learning system, and as an evaluation of factors that user consider important would make it easier to design systems that could be adopted faster in higher education setting.

The paper is organized into the following sections. Section 2 presents the research theoretical framework where mobile learning concepts have been discussed in detail. Section 3 presents the research methodology. Section 4 describes the results. The research paper conclusions as well as possible directions of future research work are presented in section 5.

## THEORETICAL FRAMEWORK

One of the most important aspects of the design process during the development of a new application is its testing and evaluation. The importance of the evaluation activity is increased in the case of an interactive system. Another important aspect of testing an interactive application is user expectations, not all of which are limited to technical capability and which are also dependent on the ways in which users might use the particular application (Stawarski & Gadd, 2010). In addition, m-Learning offers additional challenges to any evaluation methodology as outlined below:

- Varying usage contexts – Due to the highly personalized nature of learning, the context in which an m-Learning platform is used can be extremely varied, unpredictable, and subjective based on individual user experiences. As the communication and interaction are also highly personal, so also the social usage contexts vary widely. Moreover, the usage contexts are not even static for individual users but are likely to experience off-the-cuff changes, making it difficult to observe, predict, and simulate (Sharples, Milrad, & Vavoula, 2009).

- The learning process is not defined – The very idea of m-Learning blends formal and informal learning, the choice of which can be entirely made based on individual user quirks and impulses. Having merely free and easy access to learning materials does not ensure effective or even passable learning. In the absence of formal review techniques, the effectiveness of learning becomes a moot point. However, conversion to a formal activity removes the core advantage of platform flexibility (Sharples, Milrad, & Vavoula, 2009).

- Usage modes may vary – The usage of an application may be entirely different from the

intent of the designer. The personalized aspect means too many views from too many people. When the end usage is an important activity like education and knowledge gathering, the designers need to standardize the application, which is extremely difficult to do (Sharples, Milrad, & Vavoula, 2009).

- <u>Course goals and objectives</u> – While doing a performance evaluation of the m-Learning system, it is extremely important to first outline learning objectives and learner needs. The personalized nature of the platform makes it very difficult to discern the factors, as the purpose of the m-Learning platform is to make learning asynchronous, free of space constraints, and viable across different contexts (Matias & Wolf II, 2013).

- <u>Evaluation across platforms</u> – Even when the learning is formalized, the devices used to access the platform are varied. This poses significant problems to educators attempting to assess student progress and collate this with the progress of the other students. The evaluation system has to take into account the actual device used and the ease of usage in general by the particular learner. This has to be factored in while evaluating the student assessment, leading to a higher than usual technical competency requirement for educators. This means that the user dissatisfaction of minority communities cannot be neglected and it has to be taken care of platforms by the evaluation system (Matias & Wolf II, 2013).

The purpose of the present study is to discover the important factors responsible for effective m-Learning by conducting a systematic review of important and relevant studies conducted in this area by various researchers within the context of universities (higher education).

## RESEARCH METHODOLOGY

*Research Objectives*

This study attempts to answer the following main research question: *What factors are critical to the success of the m-Learning in the perspective of universities' students?* The purpose of the study is to understand if the factors leading to effective m-Learning in the higher education (university level students). The answer to the research question was gained by conducting a systematic review of the available quantitative studies in the area.

*Research Design*

The search process for this study started with a Google Scholar search with the search phrase 'critical success factors mobile learning'. The studies obtained as a result of the search were then reviewed for their data and the literature citing previous research in this field was reviewed to locate additional studies that might have similar information or related primary research data. An attempt was also made to include as many countries as possible for the research because this would give a more balanced view of the actual global status of m-Learning. As m-Learning has global ramifications – learners can be located anywhere on the globe – it makes more sense to conduct global research in the area so that the information resulting from this study could be used by universities to create a universal learning platform that would attract students from all parts of the world.

*Inclusion & Exclusion Criteria*

As m-Learning is an extremely recent concept, there was no need to exclude data that was old. In fact, the real implementation of m-Learning came after 2007 when Apple introduced the iPhone. Hence, the all-important date-based exclusion criterion was not employed; on the contrary; there was a need to capture as much primary data as possible to give a more detailed and true picture of the status of m-Learning. Our 'inclusion criteria' were related to the type of data included in the research papers:

- Research papers that used the Likert scale for assessing participant responses (regardless of the scale length)
- The complete details of the Likert scale data for the responses used in the study for each variable under assessment.

Following were the 'exclusion criteria':

- Research papers that did not use the Likert scale for assessing participant responses (i.e.,, included only percentage agreement/disagreement)
- Research papers that did not present the actual Likert scale data (for instance in several studies only the correlation/regression statistics was given that had been derived from the Likert scale data, but original data was not given)
- No research papers were included that had only qualitative data
- Duplicate reports of the same study (at least five studies were rejected on this basis). In such cases, the reports selected were those that had more primary data information and not only publication prestige
- Research studies that had only procedural information (at least two research studies belonged to this category, where the assessment procedure was cited and used in other studies but the original paper had no primary data, only the methodology)

*Quality Assessment*

The quality of each study was performed in the same way as Kitchenham's study, by employing a modified version of Database of Abstracts for Reviews and Dissemination (DARE) criteria developed by the Centre for Reviews and Dissemination (CDR) (Kitchenham et al., 2009). The original DARE criteria was used for conducting the quality assessment of systematic literature reviews. As our study uses actual original studies for analysis, the quality assessment of the studies is different even though four quality assessments similar to DARE criteria were used:

Q1. Does the research study use the 5-point Likert scale studies?
Q2. Does the research study mention the percentage of the population actually owing a mobile device and already using it for m-Learning purposes?
Q3. Does the research study divide the population based on gender?
Q4. Does the research study include responses from both students and educators?

The four questions were scored as below:

Q1. Y (Yes), there is no need for conversion and N (No), a different scale was used for assessing the responses and a conversion of the scare into the five-point scale is required. A simple formula has been used here:

$$Converted\ score = \frac{Original\ Score}{Original\ Scale} \times 5$$

- Q2. Y (Yes), complete details of participant mobile phone usage are available for this research study; N (No), absolutely no details of participant mobile phone usage are available for this research study; and P (Partly), only partial details of participants' mobile phone usage are available for this research study.
- Q3. Y (Yes), the research study divides the population specifically into male and female participants; N (No), the research study does not divide the population into male and female participants
- Q4. Y (Yes), the research study contains responses from both students and educators; N (No), the research study does not contain the responses from both students and educators.

The scoring procedure was also similar to Kitchenham: Y=1; P=0.5, N=0. Since the evaluation is based on presence or absence of information and is not qualitative in nature, the value assignment is not subjective to any individual researcher's opinion. This gives additional objectivity to the systematic nature of this study.

*Data Collection*

The data extracted from each study was divided into two segments – the collection of responses of participants and the availability of the platform to participants. The first segment is used for the derivation of the success factors and their importance to successful m-Learning implementation. The second segment can be used to assess the actual penetration of general mobile usage and the awareness of the m-Learning among the users.

Accordingly, for the first segment, CSF data, the following data was extracted:

- The source of the research study and full reference
- Author information and country where the research was actually conducted
- Population and gender distribution and user classification (students/educators or both)
- Likert scale & Actual score on the Likert scale (converted into score on a 5-point scale)
    - The individual scores for 20 individual CSFs were derived
    - Factors – discussion with students, discussion with teachers, discussion tool quality, and accessing discussion – were grouped into the CSF: learner community development.
    - Factors – hardware know-how, software know-how, browser know-how, and overall know-how – were grouped into the CSF: technical competence of students
    - After grouping, there were a total of 14 CSFs. In the absence of individual CSFs, the average of existing CSFs was taken as the scores for learner community development and technical competence of students.

For the second segment, platform availability, the following data was extracted:
- The source of the research study and full reference
- Author information and country where the research was actually conducted
- Population and how the data was presented (e.g., in percentage form or absolute numbers)
- The percentage (available or converted from absolute numbers), of users with:
    - Wireless device availability
    - Internet access
    - Access to data services, like SMS services
    - Presently using their mobile phones to access any m-Learning platform

- Interested in using their mobile phones to access m-Learning

*Data Analysis*

From the initial raw data collected from individual studies, data was tabulated systematically into multiple tables for analysis as below:
- A table measuring the quality evaluation of individual studies.
- A table measuring the Likert scale scores for the 20 CSFs, with author name, country of study, year of study, and population and gender distribution, if any.
- A table measuring the m-Learning availability, know-how, and interest among users with the number of studies and population.

From the information available, average scores were taken for the percentages for platform availability and Likert scores (converted into 5-point scale, where required). This combined with the total no. of studies that had the information (CSF weight) gives the relative importance of the CSF.

# RESULTS

The results from the systematic review are summarized and presented in this section. A total of 30 studies were eventually used in the present analysis.

*Quality Evaluation of Individual Studies*

The quality of each individual study was based on a score on the modified DARE criteria. The results of the quality assurance scores based on answers to the four quality assurance questions are shown in Table 1. None of the studies score a 4 on the quality assurance scale. This clearly demonstrates the diversity in the m-Learning assessment studies and shows that there has been no standardized assessment scheme for the studies, indicating a dire need for a standardized assessment framework in the area.

*Information on Platform availability*

From Table 2 and 3 it can be seen that out of the total of 30 studies, 12 do not have any information-reading platform availability. This means that we do not have any information about the mobile platform availability or interest in m-Learning usage for about 34.2% of the population. Researchers have inquired about the availability of mobile phones in 17 cases (3,202 population sample). It was found that an overwhelming majority, 91.63%, of the sample population in the study owned a mobile phone, which corroborates the immense penetration of mobile technology in recent times. It can be reasonably concluded that access to a mobile phone would not pose a barrier to the success of m-Learning. In 11 cases, the researchers made an inquiry into the access to Internet and access to data services like short message service (SMS). This is important information, since either of the two ways are the primary ways in which students would have access to the m-Learning content, whenever they want and wherever they are.

*Table 1. Quality evaluation of individual studies*

| ID | Author names | Q1 | Q2 | Q3 | Q4 | Total |
|---|---|---|---|---|---|---|
| S1 | (Liaw, Hatala, & Huang, 2010) | N | N | Y | N | 1 |
| S2 | (Motiwalla, 2007) | Y | P | N | N | 0.5 |
| S2A | (Motiwalla, 2007) | Y | P | N | N | 1.5 |
| S3 | (Mac Callum, 2009) | Y | P | Y | N | 2.5 |
| S4 | (Conradie, Lombard, & Moller, 2013) | Y | P | Y | N | 2.5 |
| S5 | (Alzaza & Yaakub, 2011) | Y | P | Y | N | 2.5 |
| S6 | (Ismail, Bokhare, Azizan, & Azman, 2013) | Y | P | Y | Y | 3.5 |
| S7 | (Maniar, Bennett, & Gal, 2007) | Y | N | N | N | 1 |
| S8 | (Zengning, 2011) | Y | N | Y | N | 2 |
| S9 | (Shih & Chuang, 2010) | N | N | N | N | 0 |
| S10 | (Imran, 2007) | Y | N | N | N | 1 |
| S11 | (Alzaza, 2013) | Y | Y | Y | N | 3 |
| S12 | (Huang, Yang, Huang, & Hsiao, 2010) | N | N | Y | Y | 2 |
| S13 | (Jamaldeen, Hewagamage, & Ekanayake, 2012) | Y | Y | N | N | 2 |
| S14 | (Suresh & Al-Khafaji, 2009) | Y | N | N | N | 1 |
| S15 | (Adedoja, Adelore, Egbokhare, & Oluleye, 2013) | N | N | N | N | 0 |
| S16 | (Corlett, Sharples, Bull, & Chan, 2005) | Y | N | N | N | 1 |
| S17 | (Chang, Yan, & Tseng, 2012) | N | N | Y | N | 1 |
| S18 | (Uzunboylu, Cavus, & Ercag, 2009) | Y | N | Y | N | 2 |
| S19 | (Donaldson, 2012) | N | Y | Y | N | 2 |
| S20 | (Moura & Carvalho, 2009) | N | Y | Y | N | 2 |
| S21 | (Khwaileh & AlJarrah, 2010) | Y | Y | Y | N | 3 |
| S22 | (Al-Fahad, 2009) | Y | P | Y | N | 2.5 |
| S23 | (Thornton & Houser, 2005) | N | Y | Y | N | 2 |
| S24 | (Knezek & Khaddage, 2012) | Y | Y | N | N | 2 |
| S25 | (Cheong, Lee, Crooks, & Song, 2012) | N | N | Y | N | 1 |
| S26 | (Özdoğan, Başoğlu, & Erçetin, 2012) | Y | P | Y | N | 3 |
| S27 | (Liu, Li, & Carlsson, 2010) | N | P | Y | N | 2 |
| S28 | (Scornavacca, Huff, & Marshall, 2009) | Y | Y | Y | N | 3 |
| S29 | (Liaw S.-S. H.-M., 2011) | N | N | Y | N | 1 |
| S30 | (Motiwalla, 2008) | Y | P | N | N | 2 |

*Table 2: Platform availability information for the study population*

| ID | Country | Population | Availability of Mobile Phone | Internet Access | Access to data services | Already using mobile phone for m-Learning | Interested in using mobile phone for m-Learning |
|---|---|---|---|---|---|---|---|
| S1 | China | 152 | NA | NA | NA | NA | NA |
| S2 | USA | 19 | 84.21 | 43.75 | NA | NA | 57.89 |
| S2A | USA | 44 | 86.36 | NA | 63.64 | 79.55 | 64.63 |
| S3 | New Zealand | 30 | 89 | NA | NA | NA | NA |
| S4 | South Africa | 54 | 100 | NA | 100 | 100 | NA |
| S5 | Malaysia | 261 | 95.1 | NA | 81.3 | 80.1 | NA |
| S6 | Malaysia | 38 | NA | NA | NA | 71.05 | 89.47 |
| S7 | UK | 45 | NA | NA | NA | NA | NA |
| S8 | China | 24 | NA | NA | NA | NA | NA |

| | | | | | | |
|---|---|---|---|---|---|---|
| S9 | Taiwan | 32 | NA | NA | NA | NA | NA |
| S10 | Pakistan | 438 | NA | NA | NA | NA | NA |
| S11 | Palestine | 378 | 97.4 | 69.8 | 60.3 | 79.1 | 85.2 |
| S12 | Taiwan | 147 | NA | NA | NA | NA | NA |
| S13 | Sri Lanka | 154 | 99 | 63 | 64 | 85 | 95 |
| S14 | UK | 26 | NA | NA | NA | NA | NA |
| S15 | Nigeria | 201 | NA | NA | NA | NA | NA |
| S16 | UK | 17 | NA | NA | NA | NA | NA |
| S17 | Taiwan | 158 | NA | NA | NA | NA | NA |
| S18 | North Cyprus | 41 | NA | NA | NA | NA | NA |
| S19 | USA | 330 | 95.15 | 79.1 | 84.24 | 87.27 | 86.7 |
| S20 | Portugal | 15 | 100 | 87 | 73 | 80 | 93 |
| S21 | Jordan | 314 | 86 | NA | NA | 80.32 | 80.73 |
| S22 | Saudi Arabia | 186 | 47 | 43 | 45 | 25.3 | 74.4 |
| S23 | Japan | 333 | 100 | 83 | 100 | 61 | 100 |
| S24 | USA | 81 | NA | NA | NA | NA | NA |
| S25 | USA | 177 | 86 | NA | NA | NA | 87.2 |
| S26 | Turkey | 81 | 84 | 30 | NA | NA | 80 |
| S27 | China | 209 | 93.3 | 64.59 | NA | 56 | 100 |
| S28 | New Zealand | 569 | 96.8 | 64.9 | 82.8 | 30 | 90 |
| S29 | Taiwan | 168 | NA | NA | NA | NA | NA |
| S30 | USA | 33 | 91 | 45.45 | 93.9 | NA | 75.76 |

*Table 3: Summary statistics of the Platform availability primary data*

| Mobile Platform Availability | No. of Studies out of 30 | Population | Percentage of total Population |
|---|---|---|---|
| No Information | 12 | 1626 out of Total of 4755 | 34.2% |
| Availability of Mobile Phone | 17 | 2934 out of Total of 3202 | 91.63% |
| Internet Access | 11 | 1565 out of Total of 2551 | 61.35% |
| Access to data services | 11 | 1831 out of Total of 2372 | 77.19% |
| Already using a mobile phone for m-Learning | 13 | 1855 out of Total of 2900 | 63.97% |
| Interested in using mobile phone for m-Learning | 14 | 2565 out of Total of 2915 | 88% |

Of the 1,565 sample population, about 61.35% had access to the Internet, clearly showing the lack of Internet access of a significant sample population; the cause of this could be due either to prohibitive cost or coverage issues. Similarly, of the 1,831 sample population, about 77.19% had access to data services. The reason behind this lack could be prohibitive costs and/or lack of reasonable usage plans on the part of the local mobile phone operators. This, too, could be a hindrance to the success of m-Learning. In 13 studies, researchers inquired whether students had experience in or were currently using mobile phones to access m-Learning. The results were encouraging, since of the 1,855 population about 63.97% reported having already used or currently using their mobile phones for accessing m-Learning. This number might be higher than either access to the Internet or data services, since more users were polled in this study. This shows that there is a high level of awareness and experience regarding the m-Learning. Finally, in 14 studies, researchers inquired about the interest in using the m-Learning; a majority, 88%, of the participants were interested in using m-Learning, indicating the popularity of the platform among potential users.

*Critical Success Factors from systematic review of studies*

As there are a total of 14 CSFs, they have been divided into two tables – Table 4 and Table 5, each containing scores on the Likert scale for the individual studies for 7 CSFs. NA indicates that a score for that CSF is not available.
In the Table 4, learner perceptions have been highlighted separately. This factor is essentially what users think of the m-Learning and is the actual factor that determines whether users are interested in using the platform in the future. Care has been taken to clearly show the studies that have user responses on a scale different from the standard and original 1-5 Likert scale. We have assessed 30 studies in our research from 17 countries – China (3), USA (6), New Zealand (2), South Africa (1), Malaysia (2), UK (3), Taiwan (3), Pakistan (1), Palestine (1), Sri Lanka (1), Nigeria (1), North Cyprus (1), Portugal (1), Jordan (1), Saudi Arabia (1), Japan (1), and Turkey (1).

The values collected in Tables 4 and 5 were averaged for all 30 studies. The results are summarized in Table 6. All the factors are assessed on a Likert scale of 1 to 5 (Strongly Disagree to Strongly Agree). A score higher than the average 2.5 shows that users are satisfied with the particular feature of the m-Learning that they are currently using. The most interesting aspect of this study is that all of the 14 factors mentioned below are considered to be important by the users, and they are satisfied with the particular feature as all the CSFs show a Likert scale response much higher than the average value of 2.5.

*Table 4: Likert scale responses for CSFs – Part A*

| ID | Technical Competence Students | Technical Competence Educators | Personalization | Learner Autonomy | **User Perception** | User Friendly Design | Application Working |
|---|---|---|---|---|---|---|---|
| S1 | 1.9* | NA | NA | 2.87* | **3.14*** | 3.94* | 3.42* |
| S2 | NA | NA | NA | NA | **3.71** | 2.68 | 3 |
| S2A | NA | NA | 3.7 | NA | **3.33** | NA | NA |
| S3 | 3.81 | NA | NA | NA | **3.22** | NA | NA |
| S4 | 4.3 | 4.16 | 4.18 | NA | **3.72** | 3.83 | 3.28 |
| S5 | NA | NA | NA | NA | **3.87** | NA | NA |
| S6 | NA | 3.96 | NA | NA | **4.21** | NA | NA |
| S7 | NA | NA | 3.66 | NA | **3.54** | 3.44 | 3.42 |
| S8 | 4.74 | NA | NA | 4 | **4.43** | NA | 3.78 |
| S9 | NA | NA | NA | 2.9** | **2.9*** | 3.12** | 3.14** |
| S10 | NA | NA | NA | NA | **3.6** | NA | NA |
| S11 | 3.5 | NA | NA | NA | **4.09** | NA | NA |
| S12 | NA | NA | NA | NA | **3.56*** | 3.64* | 3.58* |
| S13 | NA | NA | NA | NA | **3.84** | 4.33 | 4.03 |
| S14 | NA | NA | NA | NA | **3.16** | 3.11 | 3.08 |
| S15 | 4.91* | NA | NA | 2.59* | **3.97*** | 4.36* | 4.51* |
| S16 | 2.81 | NA | NA | 2.69 | **3.19** | 2 | 3.56 |
| S17 | 3.78* | NA | 3.77* | 3.86* | **3.64*** | 3.81* | 3.99* |
| S18 | NA | NA | 3.8 | NA | **3.87** | 3.75 | 3.9 |
| S19 | NA | NA | NA | 4.01* | **3.54*** | 3.94* | 3.74* |
| S20 | NA | NA | NA | 3.12*** | **4.35**** | 4.45*** | 4.27*** |
| S21 | 4.1 | NA | 3.99 | 3.89 | **4.46** | 4.07 | 4.04 |
| S22 | NA | NA | NA | NA | **3.68** | NA | NA |

| ID | | | | | | |
|---|---|---|---|---|---|---|
| S23 | 3.96** | NA | 3.83** | NA | **4.44**** | 3.39** | 3.94** |
| S24 | NA | NA | NA | NA | **4.33** | 4.17 | 4.23 |
| S25 | 3.44* | 3.21* | 3.71* | 3.86* | **3.57*** | 3.79* | 3.69* |
| S26 | NA | NA | 4.05 | 3.63 | **3.95** | 4.27 | 3.74 |
| S27 | 4.1* | NA | NA | 3.31* | **3.43*** | NA | NA |
| S28 | NA | NA | NA | NA | **3.67** | NA | 3.67 |
| S29 | 2.93* | NA | NA | 3.11* | **2.86*** | 4.09* | 3.53* |
| S30 | NA | NA | NA | NA | **3.58** | 3.59 | 3.06 |

\* Converted value from (1-7) scale; ** Converted value from (1-9) scale; *** Converted value from (1-3) scale

*Table 5: Likert scale responses for CSFs – Part B*

| ID | Learning Made Interesting | Assimilation with curriculum | Increased Productivity | Learner Community Development | Platform Accessibility | Internet Access | Blended learning |
|---|---|---|---|---|---|---|---|
| S1 | 3.08* | NA | NA | NA | NA | NA | NA |
| S2 | NA | 3.79 | NA | 3.52 | NA | NA | NA |
| S2A | NA | 3.64 | 3.89 | 4.05 | 4.27 | 3.8 | 375 |
| S3 | NA | NA | NA | NA | NA | NA | NA |
| S4 | 3.8 | 2.82 | 3.94 | NA | NA | NA | NA |
| S5 | NA | NA | 3.91 | 3.91 | 4.05 | 4.05 | NA |
| S6 | 4.39 | 4.17 | 4.08 | 3.27 | 4.8 | NA | 2.16 |
| S7 | 3.66 | NA | 3.28 | NA | 3.89 | NA | NA |
| S8 | NA | NA | 4.48 | NA | 4.65 | NA | 4.48 |
| S9 | 3.06** | 3.09** | 3.28** | NA | NA | NA | 3.09** |
| S10 | NA | 4.2 | 3.9 | 3.8 | 4.1 | NA | 3.9 |
| S11 | NA | 3.8 | 4 | 3.96 | 4.03 | 3.8 | NA |
| S12 | NA | 3.55* | NA | 3.96* | NA | NA | NA |
| S13 | 4.18 | 3.25 | 3.89 | 2.03 | 3.6 | 3.1 | NA |
| S14 | NA | NA | NA | NA | NA | NA | NA |
| S15 | 3.78* | NA | 2.46* | NA | NA | NA | NA |
| S16 | 3.18 | NA | 3.37 | NA | NA | NA | NA |
| S17 | 3.64* | 3.9* | 3.65* | NA | 3.95* | NA | NA |
| S18 | 4.12 | 3.87 | 4.04 | 3.87 | 3.85 | 3.8 | 4.02 |
| S19 | 3.11* | NA | 3.44* | 3.21* | 3.81* | NA | NA |
| S20 | 4.22*** | 4.67*** | 4.1*** | 4.32*** | 4.55*** | 4.55*** | 4.67*** |
| S21 | 4.08 | 4.08 | 4.28 | NA | 4.12 | NA | 3.89 |
| S22 | NA | NA | 2.44 | 2.47 | 2.55 | 1.96 | NA |
| S23 | 4.22** | 4.62** | 4.06** | NA | NA | NA | 4.61** |
| S24 | 3.98 | 4.09 | 4.26 | NA | NA | NA | 3.95 |
| S25 | 3.42* | 3.62* | 3.51* | 4.06* | 4* | 4.34* | 3.66* |
| S26 | 3.65 | NA | 3.85 | 3.39 | 3.92 | 4.41 | NA |
| S27 | NA | NA | 3.31* | NA | NA | NA | 3.34* |
| S28 | 4.04 | 2.95 | 3.76 | 4.05 | 3.83 | NA | 3.58 |
| S29 | NA | 2.94* | 2.95* | 3.89* | 3.89* | 4.17* | 2.86* |
| S30 | NA | NA | NA | 3.67 | 3.36 | NA | 3.9 |

\* Converted value from (1-7) scale; ** Converted value from (1-9) scale; *** Converted value from (1-3) scale

The first factor of interest is the user perception (in bold). This shows that users are, in general, happy with the existing m-Learning they are using and would like to continue to use the platform in the future. They perceive that the platform offers them sufficient benefits to warrant continuing usage. As this is the core assessment response, the fact that it is present in all the studies does not mean anything special. The presence of other factors and their effect on user perception is actually of more interest, after a cursory look at whether users found the overall system useful.

From the point of view of the research, an understanding of whether users thought the m-Learning system increased their productivity was considered to be of the utmost significance. This explains the presence of the factor in more than 90% of the studies. Users, on average, considered that using the m-Learning led to an increase in their efficiency and productivity. However, this does not mean that a lower percentage means that the factor is of less importance, merely that researchers did not include the factor as part of their research study. For instance, technical competence was assessed in only three studies, and logic states that educators need to be well-trained in the platform to give the maximum benefit to the students. Similarly, access to the Internet, which students consider extremely important, was evaluated in merely 31% of the studies. The results from the analysis can be used by prospective researchers to enhance their research studies and gain pertinent information regarding the performance and perception of the m-Learning within an institution.

*Table 6: Summary statistics of the Likert scale responses for the CSFs*

| CSF | Average Value | Number of studies Out of 30 | Population Out of 4755 | Percentage Population |
|---|---|---|---|---|
| Technical competence – students | 3.69 | 13 | 2215 | 46.58% |
| Technical competence – educators | 3.37 | 3 | 579 | 12.18% |
| Personalization | 3.86 | 9 | 1247 | 26.22% |
| Learner autonomy | 3.47 | 13 | 1878 | 39.5% |
| **User perception** | **3.68** | **30** | **4755** | **100%** |
| User-friendly application design | 3.69 | 21 | 2426 | 51.02% |
| Application working | 3.74 | 23 | 3171 | 66.69% |
| Learning made interesting | 3.76 | 18 | 2792 | 58.72% |
| Assimilation with curriculum | 3.73 | 17 | 3006 | 63.22% |
| Increased productivity | 3.52 | 25 | 4348 | 91.44% |
| Learner community development | 3.60 | 16 | 2893 | 60.84% |
| Platform accessibility | 4.01 | 19 | 3454 | 72.64% |
| Internet access | 3.96 | 10 | 1505 | 31.65% |
| Blended learning | 3.8 | 15 | 2516 | 52.91% |

## DISCUSSION

*Summary of Results*

Overall, our study identified 30 studies that contained primary data comprising the actual responses of the m-Learning users on how they evaluated the various aspects of the m-Learning that was tested in their institution. The study contains research conducted in 17 countries worldwide with a combined sample population of 4,755 (majority being students using m-Learning in various courses). Overall, the research showed that the users were seen to be fairly satisfied with the usage of m-Learning within their particular courses and were interested in using the system more in the future. On a 1-5 Likert scale, the satisfaction ratio was a respectable 3.68, which clearly shows a positive response.

While universal response about the availability of mobile phones and related services was not available, the studies that included this information found that more than 90% of the sample population claimed to own mobile phones. Similarly, although information regarding access to the Internet and data services was not universally available, more than 61% and 77% of the population, respectively, had access to these services. Interestingly, about 66% of the population (for the studies where information was available), had already used m-Learning platforms and an overwhelming 88% of the population was interested in using mobile phones for m-Learning purposes. It is important that future studies conducted in this area have information on these aspects, as this would give a clear picture of the actual status of m-Learning in a particular institution and of possible technological barriers that need to be overcome in individual cases.

*Discussion on critical success factors*

The information available about the CSFs is highly subjective to the individual researchers. Interestingly, all 13 factors were found to be necessary to the success of m-Learning. Even without considering the number of studies that assessed the success factors, the results of the present research can be used for indications of the relative importance of critical factors from the point of view of the users.
Platform accessibility was considered to be the most important factor, followed by Internet access, personalization of the platform, the possibility of blended learning, and the prospect of learning made interesting. This showed that the factor judged to be the most important was the involvement of the university administration in providing clear access, goals, and guides to using the platform. The second most important factor was access to the Internet, and the third most important factor was personalization of the platform. This is interesting because this shows that while students may or may not be interested in learner autonomy, they are extremely interested in the possibility of tailored learning that would satisfy individual learning goals and objectives. The next most important factor was blended learning. Users also rated the prospect of mobile phones offering an interesting way to learning to be a key success factor. This factor becomes even more important in light of the fact that m-Learning is mostly controlled at the learners' pace and time, and it would not work efficiently if users are not interested in the learning itself. These top five CSFs need to be kept in mind if a new m-Learning is to find sustainable long-term success.

The other eight success factors, in decreasing order of importance, are – application working, assimilation with curriculum, technical competence of students, user friendly application design, learner community development, increased productivity, learner autonomy, and technical competence of educators. A remarkable aspect of the results is that, while the factors are rated in the decreasing order of importance, the least important factor has a Likert score of 3.37, which is significantly higher than average. Also, all the factors are close to each other with less than the maximum distance between adjoining factors of $\leq 1$. When this information is combined with the fact that not all of the factors have been evaluated as part of all studies and that some CSFs have been evaluated as less, as three to 10 studies out of 30 show, the factors are fairly close to each other in importance and cannot be ignored in favor of others.

## CONCLUSION

This research work presents an exhaustive systematic survey of the existing research studies evaluating m-Learning worldwide. The study in particular is based on the perspective of university students. The systematic review collated the responses from 4,755 respondents collected in 30 studies conducted in 17 countries worldwide. The results of the systematic review showed that the research conducted in the area of m-Learning was fragmented and idiosyncratic and based on the understanding of the individual researcher.
A total of 13 critical success factors were evaluated as part of the study along with a measurement of user perceptions of the m-Learning. All 13 factors were found to have a significant impact on the success of the m-Learning from the user perspectives. It was also found that users were satisfied with the m-Learning and were interested in using of it in future. M-Learning was also considered to improve efficiency and productivity among the users. The future focus could be to evaluate the impact of individual success factors on the overall perception of the platform. This would quantify the effect of each success factor in precise statistical terms, and it which would be a relevant basis on which to design and implement future m-Learning.